\documentclass[prd,aps,twocolumn]{revtex4}
\usepackage{amsmath}
\usepackage{graphicx}
\usepackage{amssymb}
\usepackage{color}

\def\mp{M_{\rm p}}
\def\be{\begin{equation}}
\def\ee{\end{equation}}
\def\bea{\begin{eqnarray}}
\def\eea{\end{eqnarray}}
\def\be{\begin{equation}}
\def\ee{\end{equation}}
\def\ba{\begin{eqnarray}}
\def\ea{\end{eqnarray}}

\def\d{{\rm d}}

\newcommand\lsim{\mathrel{\rlap{\lower4pt\hbox{\hskip0.5pt$\sim$}}
    \raise1pt\hbox{$<$}}}
\newcommand\gsim{\mathrel{\rlap{\lower4pt\hbox{\hskip0.5pt$\sim$}}
    \raise1pt\hbox{$>$}}}

\begin{document}

%
%
\input epsf
\renewcommand{\topfraction}{0.99}
\renewcommand{\bottomfraction}{0.99}

\title{Enhancing Inflationary Tensor Modes  through Spectator Fields}
\author{Matteo Biagetti{$^{a}$}, Matteo Fasiello{$^{b}$}  and Antonio Riotto{$^{a}$}}
\address{$^{a}$ D\'epartement de Physique Th\'eorique and Centre for
Astroparticle Physics (CAP),\\ \textcolor{white}{} Universit\'e de Gen\`eve, 24 quai E. Ansermet, CH-1211 Gen\`eve, Suisse\\ 
$^{b}$ Department of Physics, Case Western Reserve University, Cleveland, OH, 44106}
\date{\today}

\begin{abstract}
\noindent We show that a large contribution to tensor modes during inflation  can be generated by  a spectator scalar field with speed of sound lower than unity.

\end{abstract}


\maketitle
\noindent
\noindent
The existence of gravitational waves (GW's)  is a robust prediction of general relativity \cite{maggiore}. Several detectors varying in range and sensitivity currently exist  which are probing the sky in search for such a signal. However hard the detection might prove to be, there is no shortage in the literature of possible mechanisms as to how to generate gravitational waves. The timing might also vary: in the early universe for example, GW's could have been produced by bubble collisions during first order phase transitions \cite{Kos1,Kos2}, by cosmic defects \cite{vil,dani}, as well as  after inflation, during the reheating stage \cite{Khlebnikov:1997di}. Quite in general, the study of gravitational waves represents a formidable window on the physics of early universe.

In this paper we focus on the tensor modes  production during inflation. It is well known that during inflation scalar and tensor modes are generated by initial quantum fluctuations at small scales at the linear level \cite{lrreview}. In particular, tensor modes are generated with an amplitude proportional to the Hubble rate $H$ during inflation and they  
might show up in the B-mode polarization of the cosmic microwave background radiation anisotropies \cite{pol1,pol2} which is one of the main targets of the Planck mission \cite{planckmission}. Within single-field models of inflation, there is a well-defined prediction for the tensor-to-scalar ratio, $r\sim\epsilon$, where $\epsilon$, defined by the relation $\dot H=-\epsilon H^2$,  is one of the slow-roll parameters. 

Nevertheless, and most likely, the inflaton field was not the only scalar
field dynamically excited during the inflationary stage. On the contrary, when rooted in high energy physics models, active scalar fields are ubiquitous during inflation, the only condition being that their mass is smaller than the Hubble rate $H$ during the inflationary stage.
As examples one might think of supergravity and (super)string models where there are a large number of moduli fields;  another example is to be found in the theories of extra-dimensions where Kaluza-Klein excitations appear.

In this paper we are concerned with what is called a spectator field $\sigma$, that is, a light field that plays no role whatsoever in the inflationary dynamics, neither at the background level (it does not give a significant contribution to the energy density during or after inflation) nor at the level of generating the final curvature perturbations (which we assume to be predominantly sourced by the inflaton field). 
In principle, whenever there are two or more fields one should worry, alongside adiabatic fluctuations, also about isocurvature perturbations which have been shown in some cases to contribute to the power spectrum. However, isocurvature modes survive only if the extra degrees of freedom (other than the inflaton) are stable or if they decay very late, e.g. when the baryon number is well defined. In all that follows, the spectator role of $\sigma$ is that of a field which does not in any way drive the inflationary dynamics, not the background, nor the fluctuations.

The only non-trivial property  we are assuming is that the spectator field has a speed of sound, $c_s$, smaller than unity. The corresponding action for fluctuations reads

\bea
S_{\delta \sigma}= \int \d^3x\,\d\tau a^4\left[\frac{1}{2a^2} \left( \delta \sigma{'^2} - c_s^2 (\nabla \delta \sigma)^2\right)- V_{(2)} \right], \label{mainsigma}
\eea
where $a(\tau)$ is the scale factor in terms of the conformal time $\tau$ and the superscript $'$ stands for derivative with respect to it.
An action which describes a theory with $c_s <1$ such as this one is a familiar sight within the so-called  $P(X,\phi)$ theories \cite{kinf,Chen:2006nt}, with $\phi$ being a scalar degree of freedom and where 
$X=(1/2)g^{\mu\nu}\partial_\mu\sigma\partial_\nu\sigma$, and $c_s=\partial_XP/(\partial_XP+2X\partial_{XX}^2P)$. Let us give another example of how the condition $c_s<1$ might be implemented. Consider a term such as
\bea
\Big[(g^{\mu\nu} +\partial^{\mu}\phi\,\partial^{\nu}\phi) /(1+ \partial_{\mu}\phi \partial^{\mu}\phi) \Big] \cdot  \partial_{\mu}\sigma \partial_{\nu}\sigma \,\, ;
\eea
 this non-minimal coupling also generates for $\sigma$ an effective $g^{\mu\nu}$ which supports  $c_s\not=1$. For other, similar, realizations see \cite{gelaton}. 

The precise value of $c_s$ depends then on the specific mechanism generating it. One can, however, provide some comments on general grounds. Whenever a signal is enhanced by a small speed of sound, observations do of course provide an immediate lower bound on $c_s$. From a more formal perspective, one should be aware that a very small $c_s$, and therefore an almost null gradient term in (Eq.~\ref{mainsigma}), implies, within an effective field theory approach, that more terms in the expansion are needed (a new, non-linear dispersion relation would also be a consequence) for a gradient-like term to appear.

In models where the $c_s<1$ requirement is on the inflaton side ($P(X,\phi)$, effective theories of inflation \cite{Cheung:2007st}, etc..) a $c_s$ close to zero also raises concerns as to the validity of perturbation theory \cite{Leblond:2008gg}. This is because interactions with increasingly higher space derivatives (each additional derivative comes with $c_s$ at the denominator which competes with the $H/M$ expansion parameter, $M$ being the mass of the underlying theory) would become the leading term, to the detriment of the perturbative expansion. As we shall see, in order to obtain $\sigma$-sourced tensor-to-scalar ratio which is comparable to the usual one, it suffices to require $c_s< 10^{-2}$, a condition which does not force us into the perilous terrains of the region in the theory parameters space highlighted above.
  
Upon solving the equation of motion in Eq.~(\ref{mainsigma}), one ought to normalize the wavefunction according to the prescription that guarantees the usual commutation relations to hold also in curved space time (cfr \cite{bd}). This will univocally fix the $k, c_s$ exponent.  The properly normalized wavefunctions of the $\sigma_k$ are given by

\bea
\delta\sigma_k= \frac{H}{\left(k c_s\right)^{3/2}}\left(1+ i k c_s \tau \right) e^{- i k c_s \tau}\label{wave}
\eea
and  its power spectrum on super-Hubble scales during inflation is

\bea
\mathcal{P}_{\delta\sigma}= \frac{k^3}{2 \pi^2} |\delta\sigma_k|^2=\frac{1}{c_s^3}\left(\frac{H}{2\pi}\right)^2.
\eea
The assumption that the contribution to the curvature perturbation $\zeta$ mainly comes from the inflaton field amounts to imposing that 

\be
\label{xc1}
r_\sigma\zeta_\sigma\lsim \zeta_{\rm inf},
\ee
where $r_\sigma\approx\rho_\sigma/\rho_\gamma$ is the ratio between the energy density of the spectator field and the total density at the time of the decay of the field $\sigma$, which we assume to take place during the radiation phase after inflation. 

If we suppose that the potential of the spectator field is simply $V(\sigma)=m^2\sigma^2/2$, then $\zeta_\sigma\simeq (2/3)(H/2\pi c_s^{3/2}\overline\sigma)$ and $ \zeta_{\rm inf}\sim (H/2\pi\epsilon^{1/2} \mp)$, where $\overline\sigma$ is the vacuum expectation value of the spectator field and $\mp$ is the reduced Planck scale. Eq. (\ref{xc1}) imposes

\be
\label{xc2}
c_s\gtrsim \left(r_\sigma\epsilon^{1/2}\frac{\mp}{\overline{\sigma}}\right)^{2/3}.
\ee
Henceforth, we assume that the spectator field decays promptly after inflation, in other words we require $r_\sigma\lsim1$. This is a condition which is our own input here but is not particularly stringent even if $\overline{\sigma}\sim \mp$. At the same time, this assumption will also  also suppress
the non-Gaussianity induced by the  spectator field 

\be
{\rm NG}\sim \frac{\langle\zeta\zeta\zeta\rangle}{\langle\zeta\zeta\rangle^2}\sim r_\sigma^3\frac{\mathcal{P}^2_{\delta\sigma/\overline{\sigma}}}{\mathcal{P}^2_{\rm inf}}\sim\frac{r_\sigma^3\epsilon^{2}}{c_s^6}\left(\frac{\mp}{\overline{\sigma}}\right)^4\lesssim \frac{1}{r_\sigma},
\ee
where in deriving the last relation we have used the condition (\ref{xc2}).
Despite the passive role as far as the scalar modes are concerned, the spectator field might have a role in generating tensor modes at second-order if $c_s\ll 1$. 

The contribution from the energy-momentum tensor that is not projected away when looking for the source of gravitational waves (transverse, traceless)  is proportional to $c_s^2(\nabla \sigma)^2$.
The equation of motion for the tensor modes, sourced by the scalars at second-order, then reads 
\begin{equation}
h^{\prime\prime}_{ij}+2{\cal H}h'_{ij}-\nabla^2 h_{ij}=-4\frac{c_s^2}{\mp^2}
\hat{\cal T}_{ij}^{\,\,\ell m} \partial_\ell\delta\sigma\partial_m\delta\sigma,
\label{uno}
\end{equation}
where $\mathcal{H}= a'/a$ and   $\hat{\cal T}_{ij}^{\,\,\ell m}$ is the standard projector operator
expressed in terms of the polarization vectors $e_{i}(\vec{k})$ and $\overline{e}_i(\vec{k})$ orthogonal to $\vec{k}$

\begin{eqnarray}
\hat{\cal T}_{ij}^{\,\,\ell m} &=&\sum_{\lambda=\pm}\int\frac{\d^3 k}{(2\pi)^3} e^{i \vec{k}\cdot\vec{x}}
\,e_{ij}^\lambda(\vec{k})e^{\lambda\ell m}(\vec{k}),\nonumber\\
e_{ij}^+(\vec{k})&=&\frac{1}{\sqrt{2}}\left(e_i(\vec{k})e_j(\vec{k})+\overline{e}_i(\vec{k})\overline{e}_j(\vec{k})\right),\nonumber\\
e_{ij}^-(\vec{k})&=&\frac{1}{\sqrt{2}}\left(e_i(\vec{k})\overline{e}_j(\vec{k})+\overline{e}_i(\vec{k}){e}_j(\vec{k})\right).
\end{eqnarray}
The point is that, despite one has to pay a second-order effect, the contribution to the
tensor modes is inversely proportional to (some power of) $c_s$. This happens because the spectator field fluctuations
freeze-in at a comoving length scale $k^{-1}\sim c_s\tau$ which is much smaller  than the comoving Hubble radius. Notice that the generation of tensor modes at second-order either through the curvature perturbation themselves \cite{a1,a2} or through the  curvaton field \cite{a3,a4} has been studied in the literature. 
 
First, we look at the classical solution by using the appropriate Green's function $g_k(\tau, \tau')$ for the equation
\bea
g''_k+2\mathcal{H} g'_k +k^2 g_k=\delta(\tau-\tau').
\eea
The solution takes the form
\begin{eqnarray}
g_k(\tau, \tau')&=&\frac{1}{2 k^3 \tau^{\prime\,2}}e^{-i k (\tau+\tau')} \Big[ e^{2 i k \tau} (1-i k \tau) (-i+k \tau') \nonumber \\
   &+&e^{2 i k \tau'} (1+i k \tau) (i+k \tau')\Big] \theta (\tau-\tau') , \qquad 
\eea
where for $a(\tau)=-1/H\tau$ we have assumed  the usual inflationary behavior. The general solution for $h_k(\tau)$ is then
\bea
h_k(\tau)=\frac{1}{a(\tau)} \int^{\tau} \d \tau' a(\tau') g_k(\tau,\tau') {\cal S}_{\vec k}(\tau'),
\eea
with 
\bea
{\cal S}_{\vec{k}}(\tau)= \frac{4c_s^2}{\mp^2} \sum_{\lambda=\pm}\int\frac{\d^3 p}{(2\pi)^{3}}
e^{\lambda \,\ell m}(\vec k)\,p_\ell p_m\delta\sigma_p\delta\sigma_{|\vec{k}-\vec{p}|},
\eea
${\cal S}_{\vec{k}}(\tau)$ being the Fourier transform of the source.
We now explicitly write the source in Fourier space and proceed to quantization. What we are after is the tensor power spectrum
\bea
\mathcal{P}_h= \frac{k^3}{2 \pi^2} \sum_{\lambda=\pm}|h_k|^2.
\eea
Proceeding in the standard fashion, one obtains
\bea
\mathcal{P}_h&=& \frac{32}{\pi} \frac{c_s^4}{\mp^4} \frac{ k^3}{a^2(\tau)} \int^{\infty}_{0} \d p_{}\int^{1}_{-1} \d \cos\theta \,\,p_{}^6 \sin^4\theta  \nonumber \\ 
             &\times &\left| \int_{-\infty}^{\tau}\d\tau' a(\tau') g_{k}(\tau,\tau') \sigma_{p_{}}(\tau') \sigma_{|\vec k -\vec p_{} |} (\tau')   \right|^2. \label{corta}
\eea
We perform first the $\tau$ integral below and then tackle the calculation over the momentum. The conformal time integral can be performed analytically, we give the explicit result just below. The change of variables $p \equiv k y/c_s;\,\,\, |\vec k -\vec p| \equiv k x /c_s$ has been employed,
\bea
\mathcal{P}_h &=&     \frac{ 2H^4}{ c_s^4\mp^4 }     \int_{{\small \frac{1+c_s}{2}}}^\infty \text{d}y\int _{|y-c_s|}^{y+c_s}\d x\,\, \frac{y^2}{x^2} \left(1-\frac{\left(c_s^2-x^2+y^2\right)^2}{4\, c_s^2\, y^2}\right)^2  \nonumber \\ 
&& \Bigg| e^{-i z} \left(1+x^2+y^2\right) (-i+z) \text{Ei}\left[-i (-1+x+y) z\right] \nonumber \\
&&+ e^{-i (x+y) z}\cdot \Biggl(\frac{-2 i (x+y) \left(-1+x^2+y^2\right)-4 x y z}{(-1+x+y) (1+x+y)} \nonumber \\
&&+  e^{i (1+x+y) z} \left(1+x^2+y^2\right) (i+z)  \nonumber\\ 
&& \times\,\,  \text{Ei}\left[-i (1+x+y) z\right]\Biggr) \Bigg|^2\times \frac{1}{\pi}. \label{lunga} \nonumber \\
\eea
Note that, after the change of variables, the $c_s$ dependence inside the integral is simply encoded within just one factor, it is a contribution originating from the angular $\sin^4\theta$ term. The time dependence is accounted for by the variable $z\equiv k \tau.$ 
Schematically, we can write Eq. (\ref{lunga}) as
\bea
\mathcal{P}_h = \frac{ H^4}{c_s^4\mp^4} \int_{{\small \frac{1+c_s}{2}}}^\infty\d y\int _{|y-c_s|}^{y+c_s}\d x\, \frac{F(x,y,z,c_s)}{\text{  }( x+y-1)} .
\label{pnotyes}
\eea 
This simple compact expression shall suffice for our upcoming discussion on the momentum integral. The function $F$ encodes the $\tau$-integral result and should be thought of as a well behaved function in the $(x,y)$ plane so that all the focus is on the denominator. In previous analysis, see for instance  Refs. \cite{a1,a2,a3}, the sound speed  $c_s$ was equal to unity,  meaning  that the set of $(x,y)$ points for which the divergence in the denominator played a role was zero measure and therefore unimportant. In our case  things are quite different, as an entire interval in the plane makes the denominator vanish. Indeed,  the general expression for the denominator above is
\bea
x+y-1= \frac{1}{|\vec{k}|}\Big[c_s \left( |\vec p|+|\vec p-\vec k|\right) - |\vec k|\Big]
\eea
 and,   when $c_s <1$, it becomes possible to cross the $x+y=1$ line. This fact has a simple physical interpretation that will help us handle the integrals in Eq.~(\ref{lunga}).  

Our setup consists of two types of particles, gravitons and the spectator fluctuations,  with different speeds of sound. It is a known effect that, given the appropriate interaction,  the slower particle can radiate (think of Cherenkov radiation).  In our case what takes place is that the graviton is sourced by the $\sigma$-particles. That is why the resonances show up in what is basically the graviton two-point function.
This $\sigma $-radiation could be treated as it is normally done with soft photons: by summing up all their loop contributions and showing the result is finite.

One might also take a different approach, by making use of a natural cutoff for these modes; quite simply, the suggestion comes from the $c_s = 1$ discussion above: one simply needs to require  $c_s p + c_s |\vec k-\vec p|\geq k$. This condition is clearly automatically satisfied if $c_s=1$, it is indeed the Cauchy-Schwarz inequality in that case.
One could in principle later calculate the exact effect of these resonances but, considering we are interested in the graviton two-point function, it is expected that the $\sigma$-radiation is resummed into a finite, small contribution. In what follows we  opt for the ``hard cut off'' solution. 
  One might object that, in forbidding access to the region complementary to $c_s p + c_s |\vec k-\vec p|\geq k$ we are not simply disposing of the resonance, but also of the small $(x,y)$ region where the resonance itself is not present. It is straightforward to show (we verified this numerically for the whole interval of $c_s$ values) that the contribution of that region represents a small correction to the overall numerical result. 
 \vspace{.3cm}

\begin{figure}[htbp]
\begin{center}
\includegraphics[scale=0.4]{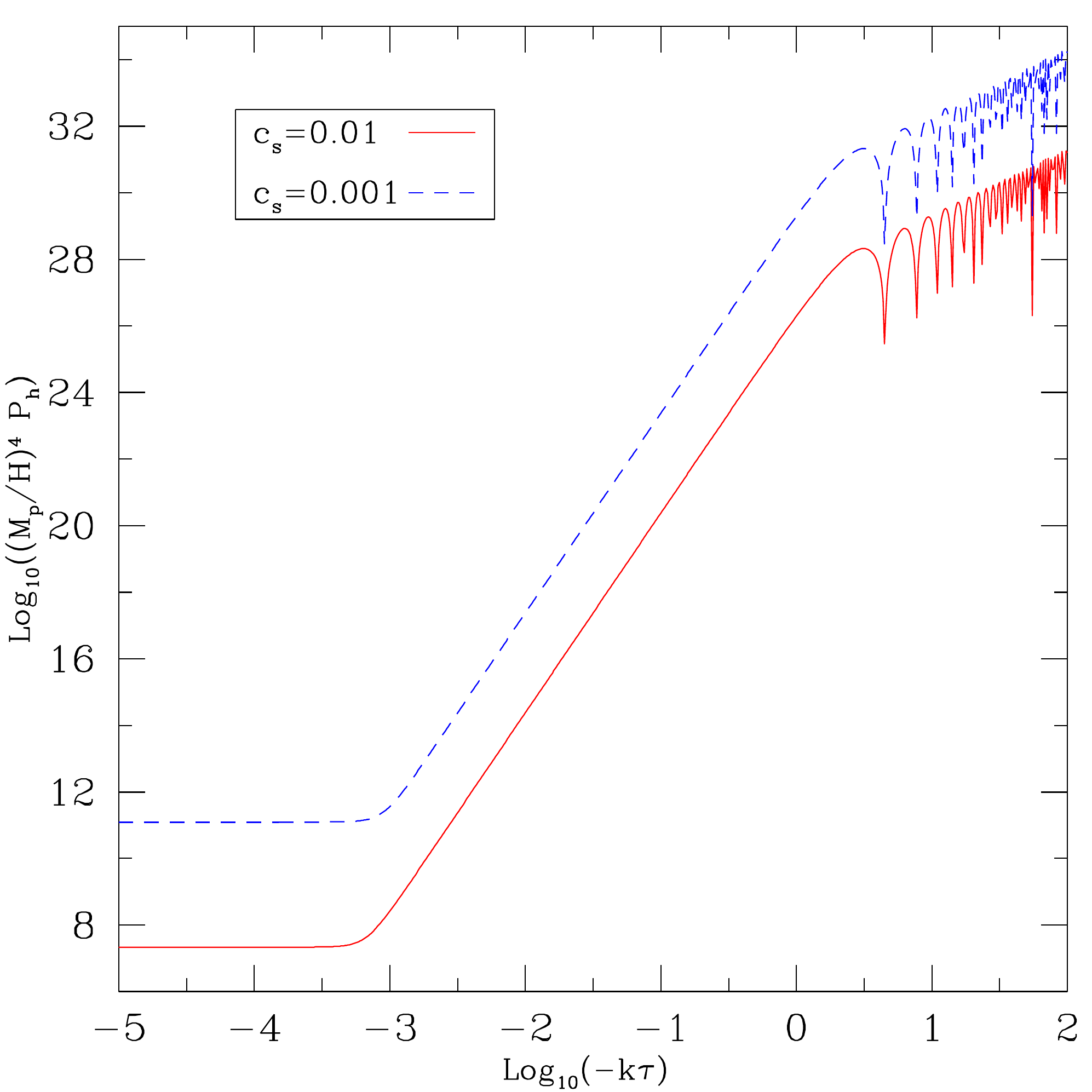}
\caption{The normalized tensor mode power spectrum  as a function of $(-k\tau)$. Either $k$ or $\tau$ can be kept fixed, both choices will generate the behavior above. }
\label{ph}
\end{center}
\end{figure}  
 Another fact which is suggestive of this resonance interpretation is that the appearance of the spike in the integral can be controlled by varying the lower $\tau$ integration  extremum. If we set the generic lower bound (one eventually sets it to be $-\infty$, projecting onto the vacuum via the $(1-i \epsilon)$ factor) to be $-M$, as $M$ increases, we see the appearance with time of the spike. This suggests it takes time for the effect to build up, as one would expect.
 \vspace{.3cm}

Having discussed above how to deal with the analytical $\tau$ integration we now take on the rest numerically, see Fig. \ref{ph}. What we are after is the $c_{s}^{-n}$ dependence, something which is ultimately responsible for the enhancement effect on the gravitational waves spectrum. Naively, from simple counting arguments one would conclude that the enhancement is of order $1/c_s^4$. Performing the numerical integration slightly modifies this educated guess: we have found that, to a good approximation, $n\simeq 18/5$, see Fig. \ref{fit}.

\begin{figure}[htbp]
\begin{center}
\includegraphics[scale=0.4]{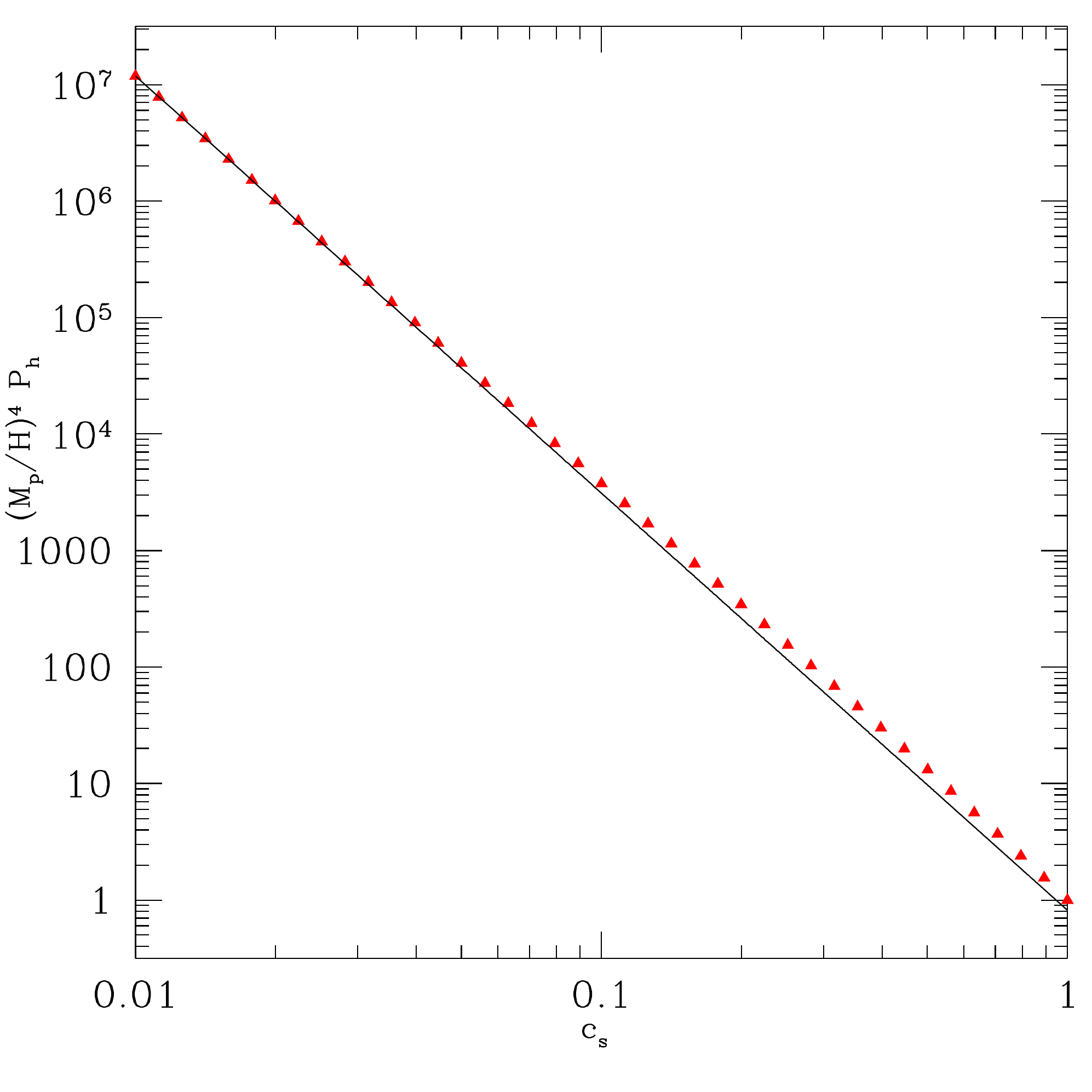}
\caption{The normalized tensor mode power spectrum as a function of $c_s$ on super-Hubble scales (red triangles) with superimposed the function with scaling $c_s^{-18/5}$.}
\label{fit}
\end{center}
\end{figure}
\noindent
 It is also useful to point out that, once all variables are taken to be dimensionless, the $k$-dependence disappears, as it should.
The final power spectrum of tensor modes on super-Hubble scales generated by the spectator field is nearly scale-invariant and its amplitude is
\bea
\label{h}
{\cal P}_{h}=   c\frac{H^4}{c_s^{18/5}\mp^4}, \label{power}
\eea
where a careful inspection of the numerical results reveals that the numerical factor is   $ c\simeq 3$. 
Notice also that this result might change if more than one spectator field is present; roughly, one would have to multiply by the number
of fields. For instance, if the spectator fields belong to a O($N$)-multiplet and such a symmetry is not broken,  one should multiply the result (\ref{h}) by a factor $N$ if all the corresponding sound speeds are of the same order of magnitude (otherwise the dominant contribution would come from the smallest of the sound speeds).

We pause here to comment more on the time-dependence of our result and on the precise dynamics of the generation of the signal. Up until Eq.~(\ref{pnotyes}) the time variable is $z$ which accounts for the full analytical temporal dependence in our setup. Upon performing the numerical integration one realizes that the signal is sourced up to horizon (Hubble radius) exit, with by far the leading contribution being at horizon crossing. Afterwards, the signal reaches a plateau so that, if one is after the late-time result, it shall suffice to consider any late time $z$ and  the time dependence automatically drops out, just as in Eq.~(\ref{power}). Hints of this behavior are already visible in Fig.1 with an inside-the-horizon oscillating profile and an outside plateau. Performing the full calculation confirms the expectations.

One can also calculate the ratio $(T_\sigma/S)$ between the tensor mode power spectrum sourced by the spectator field and the curvature perturbation power spectrum ${\cal P}_\zeta=(H^2/2\mp^2\epsilon)$

\be
\frac{T_\sigma}{S}=\frac{{\cal P}_h}{{\cal P}_\zeta}= \frac{c \,\epsilon H^2}{2\, c_s^{18/5}\mp^2}.
\ee
This ratio is larger than the standard tensor-to-scalar ratio $T/S=16 \epsilon$ in single-field models of inflation, where the tensor modes are generated at the linear level, if $c_s\lsim (H/\mp)^{5/9}\lsim  10^{-2}$ where we have used the upper bound $H\lsim 10^{14}$ GeV set by the Planck data on the Hubble rate during inflation \cite{planck}. 

One might also want to see how the tilt of  the perturbation of the spectator field propagates (quadratically) onto the tensor mode spectrum. So far we have analyzed the generation of tensor modes in a de Sitter set-up with a massless spectator field.  Taking into account that inflation takes place during a  quasi de Sitter stage and that the spectator field has  a tiny mass $m$, it is easy to convince oneself that  the tensor modes power spectrum  
receives a small tilt 

\bea
n_{T} \sim  2\left(\frac{2 m^2}{3H^2}-2 \epsilon\right)-\frac{18}{5}\frac{\dot{c}_s}{H c_s},
\eea
where we have taken into account that the sound speed might also slowly vary during inflation. The spectrum can therefore be blue, contrary to the spectral index from the  standard contribution which is always red, $n_T=-2\epsilon$. 

Our results might be soon relevant as the next Planck data release (see also \cite{Chiang:2009xsa,EssingerHileman:2010hh,Kogut:2011xw,Andre:2013afa} ), about one year from now,  is supposed to focus on the B-mode polarization of the cosmic microwave background anisotropies and this observable might turn out to be a way of testing, albeit indirectly,  the presence of spectator fields during the inflationary epoch. For a recent study of different mechanisms of gravitational waves production during inflation see \cite{Carney:2012pk,Barnaby:2012xt}.

\vskip 0.5cm

\noindent We would like to thank D. Figueroa for a careful reading of the draft and A.J.~Tolley for many illuminating discussions. A.R. is supported by the Swiss National
Science Foundation (SNSF), project `The non-Gaussian Universe" (project number: 200021140236), M.B. is also supported by the Swiss National Science Foundation (SNSF).



\end{document}